\makeatletter
\def\setcaptype#1{\def\@captype{#1}}
\makeatother
\newcommand{\be}{\begin{equation}}
\newcommand{\ee}{\end{equation}}
\newcommand{\bea}{\begin{eqnarray}}
\newcommand{\eea}{\end{eqnarray}}

\documentstyle[11pt,epsf]{article}
\title{Reanalysis of hyperon beta decay data on $F/D$}

\author{B.~Ehrnsperger and A.~Sch\"afer
\\
Institut f\"ur Theoretische Physik, J.~W.~Goethe
Universit\"at Frankfurt,
\\
Postfach~11~19~32, D-60054~Frankfurt am Main, Germany
\\
UFTP preprint 377-1994, hep-ph/9411267 }
\begin{document}
\maketitle
\begin{abstract}
We reanalyse hyperon beta decay data to extract the corresponding
F and D values. We show that flavour symmetry breaking effects
lead naturally to a reduction of the $F/D$ ratio.
For proton and neutron our analyses suggests $F/D = 0.49 \pm 0.08$ instead of
$F/D = 0.575 \pm 0.016$ which is generally  used to analyse the spin data.
Our smaller $F/D$ value allows to fullfil the Ellis--Jaffe sum rule
implying an unpolarized $s\bar s$-sea.
\end{abstract}
\\ \\
The original EMC data on the spindependent structure function $g_1(x)$ of the
proton \cite{JAs88} has generated great excitement as its analysis suggested
that
quarks carry only a small fraction of the nucleon spin and that the strange
quark
sea is substantially polarized in the direction opposite to the nucleon spin.
Recent theoretical investigations  and improved experimental results
\cite{BAd94}
have reduced the size of the effect but at the same time have increased the
precision such that the statistical significance of e.g. the strange quark
polarisation is still about 2 $\sigma$. The total spin carried by the quarks
is about 40 percent according to these data.

Already at the very beginning it was argued, however, that the validity of
the flavour-SU(3) symmetry used in deriving these far reaching conclusions
might be questionable \cite{HLi89,ASc88}.
It should be kept in mind that a
reduction of the $F/D$ value usually used by just 15 percent would be sufficent
to bring the data in agreement with the assumption of an unpolarized
strange-quark sea, and flavour-SU(3) violating effects are often of this
magnitude. In this contribution we want to sharpen these concerns, arguing that
taking flavour-symmetry violation into account tends to reduce the $F/D$ value.

It is well known that
semileptonic weak decay data and the spin structure functions of the nucleons
test both the axial-vectorial matrix elements. The main difference is that
the weak interaction connects different states in the baryon octet, while the
structure functions are related to matrix elements which are diagonal in
flavour space.

\begin{figure}[htb]
\center{\hspace*{0mm} \epsfxsize=17.cm \epsfysize6cm
        \epsfbox{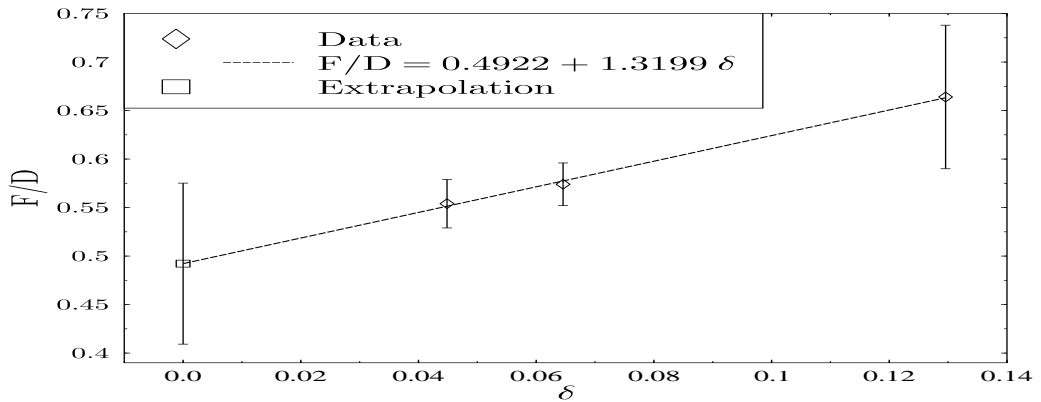}}
\caption{\label{fig1}
F/D ratios from the decays
$\Lambda \to $ p; $\sum^- \to $ n; $\Xi^-  \to \Lambda $
with the constraint F + D = $g_A/g_V (n \to p) =
1.2573 \pm 0.0028 $.}
\end{figure}

As already noticed earlier \cite{HLi94}
in the presence of flavour symmetry breaking
it is imposssible to disentangle
the information on the nucleon and the hyperons without
some definite hyperon model.
Instead of constructing such a model, we just assume that the
mass differences are a measure for the flavour SU(3) symmetry breaking.
Thus it is natural to assume that the symmetry breaking effects for $F/D$
are proportional to some dimensionless function of the mass differences.
The simplest form for such a function is
\be
\delta = \frac{(m_i + m_f) - (m_p + m_n)}{(m_i + m_f) + (m_p + m_n)}
\ee
with $m_i$ , $m_f $ ,$m_p $ and $ m_n $ , denoting the masses
of initial hadron, final hadron, proton and neutron.
$\delta = 0$ corresponds to the F and D values of proton and
neutron (we assume the validity of isopin symmetry).
If all baryons in the octet had the same mass $\delta$ would be identically
zero;  introducing the physical masses it just scales with the mass
difference between
the two states involved and twice the averaged nucleon mass.
Table 1 and figure \ref{fig1} show the F/D ratios from the decays
$\Lambda \to $ p; $\sum^- \to $ n; $\Xi^-  \to \Lambda $
with the constraint F + D = $g_A/g_V (n \to p) =
1.2573 \pm 0.0028 $.

To get these values we assumed SU(3) symmetry and
took $g_A/g_V$ from \cite{PaDa94}.

\renewcommand{\arraystretch}{1.3}
\begin{center}
\begin{tabular}{|c|c|c|c|}
\hline
decay & F/D & $\delta $ & $g_A/g_V  = \dots $ \\
\hline
 n $\to $ p            & 0.492 $\pm 0.083 $ & 0
& $ 1.2573 \pm 0.0028 = F + D $  \\
\hline
 $\Lambda \to $ p      & 0.554 $\pm 0.025 $ & 4.485 $\cdot 10^{-2}$
& $0.715 \pm 0.015 = F + D/3$ \\
\hline
 $\sum^- \to $ n       & 0.574 $\pm 0.022 $ & 6.457 $\cdot 10^{-2}$
& $-0.340 \pm 0.017 = F - D $ \\
\hline
 $\Xi^-  \to \Lambda $ & 0.664 $\pm 0.074 $ & 1.296 $\cdot 10^{-1}$
& $0.25 \pm 0.05 = F - D/3 $ \\
\hline
\end{tabular}

Table 1.
\end{center}

Also shown is the extrapolation to $\delta = 0$ ,
which corresponds to the $F/D$ ratio for proton and neutron,
with $F/D = 0.49 \pm 0.08 $.
To obtain this extrapolation (and the corresponding errors)
we assumed a linear correlation between $F/D$ and $\delta $, which
is certainly justified as long as $\delta$ is small.

The precise value of $F/D$  extrapolated to the nucleon states depends
somewhat on the definition of $\delta$. If the denominator would be
substituted
e.g. by some constant mass $M = m_p + m_n$ the result is
$F/D = 0.50 \pm 0.08$ but it should be obvious
from figure 1 that in any case one can get a value substantially reduced
with respect to that of previous
data analysises \cite{FCl93,HLi94}.
To summarize, the two main effects of the assumed correlation are:

1.) The error for F/D$_{\rm proton}$ is much larger than just the
    weighted average of the measurements.

2.) Data seem to indicate a positive correlation between
    $\delta $ and F/D.
    This results in a F/D$_{\rm proton}$ ratio substantially
    smaller than the average of all measurements.

Let us discuss next the consequences of these results for the polarized
structure functions.
The Bjorken sum rule \cite{JBj66},
which is a rock-solid prediction of QCD,
is unaffected by our data analysis.
The Ellis--Jaffe sum rule \cite{JEl74},
which is not related to any fundamental symmetry,  is, however,
strongly affected .

With the original Ellis--Jaffe asumption $\Delta s = 0$
one gets ($I_p (Q^2) = \int_0^1 dx g_1^p (x,Q^2)$ etc.):
\be \label{eq2}
I_p(Q^2) = \frac{3F + D}{18}C_{NS}(Q^2)
      + \frac{3F - D}{9}C_{S}(Q^2)
      - 0.018 \frac{GeV^2}{Q^2}
\ee

\be \label{eq3}
I_n(Q^2) = -\frac{D}{9} C_{NS}(Q^2)
      + \frac{3F - D}{9}C_{S}(Q^2)
\ee

\be \label{eq4}
I_d(Q^2) = \frac{3F - D}{36}C_{NS}(Q^2)
      + \frac{3F - D}{9}C_{S}(Q^2)
      - 0.009 \frac{GeV^2}{Q^2}
\ee
with the $Q^2$ dependence of \cite{SLa91}:
\be
C_{NS}(Q^2) = 1 - \frac{\alpha_s(Q^2)}{\pi} -
            3.5833\left(\frac{\alpha_s(Q^2)}{\pi}\right)^2
            - 20.2153\left(\frac{\alpha_s(Q^2)}{\pi}\right)^3
\ee

\be
C_{S}(Q^2) = 1 - \frac{\alpha_s(Q^2)}{3\pi}
               - 0.5496 \left(\frac{\alpha_s(Q^2)}{\pi}\right)^2
\ee
The higher twist contribution in equations
(\ref{eq2}) - (\ref{eq4}) is based on the QCD sum rule estimates
\cite{IBa90,ESt94}, the explicit form used here can be found in \cite{BEh94}.
Table 2 shows the resulting values
with $\alpha_s(Q^2)$ from \cite{PaDa94}
and the corresponding experimental data (lower row).

\begin{center}
\begin{tabular}{|c|c|c|c|}
\hline
I$_p(Q^2 = 3 $ GeV$^2$) & I$_p(Q^2 = 10 $ GeV$^2$)
  & I$_n(Q^2 = 2 $ GeV$^2$) & I$_d(Q^2 = 4.6 $ GeV$^2$)\\
\hline
0.134 $\pm $ 0.025 & 0.145 $\pm $ 0.025
           & -0.034 $\pm $ 0.025 & 0.051 $\pm $ 0.025 \\
\hline
0.129 $\pm $ 0.004 $\pm $ 0.009 & 0.136 $\pm $ 0.011 $\pm $0.011
           & -0.022 $\pm $0.011 & 0.023 $\pm $0.020 $\pm $0.015 \\
ref. \cite{KAb94} & ref. \cite{DAd94} & ref. \cite{PAn93}
& ref. \cite{BAd93} \\
\hline
\end{tabular}

Table 2.
\end{center}

The nice agreement between the Ellis--Jaffe predictions
and the various experiments leads us to the following statement:
{}From the experimental data one can either conclude
that the (strange) sea is strongly polarized (and flavour SU(3) is perfect),
which is the standard conclusion,
or that the flavour SU(3) is slightly broken
and the sea is unpolarized.
Since the second conclusion is much less spectacular it appears to
us as more probable.
In \cite{BEh94a} we showed, that it is in fact possible to fit
the Bjorken $x$ dependent polarized structure functions without sea
polarization for a phenomenological model which includes explicite
flavour SU(3) breaking.
Thus we conclude that from the present data on polarized structure functions
one cannot derive that the strange quark sea is strongly negatively polarized.

This work was supported in part by DFG (G. Hess Program) and Cusanuswerk.
\enspace A.S. thanks also the MPI f\"ur Kernphysik in Heidelberg for its
hospitality.

\newpage

{\LARGE \bf Figure Caption}

\rm
Figure \ref{fig1}:
F/D ratios from the decays
$\Lambda \to $ p; $\sum^- \to $ n; $\Xi^-  \to \Lambda $
with the constraint F + D = $g_A/g_V (n \to p) =
1.2573 \pm 0.0028 $.
\end{document}